\documentclass[prb,aps,twocolumn,floatfix,showpacs,preprintnumbers,amsmath,amssymb,10pt,letterpaper]{revtex4-1}
\usepackage{amsmath}
\usepackage{tabularx}
\usepackage{booktabs}
\usepackage{verbatim}
\usepackage{graphicx}
\usepackage{dcolumn}
\usepackage{bm}
\begin{document}


\title{Spin Accumulation at Ferromagnet/Non-magnetic Material Interfaces} 
\author{ Matthew R. Sears } 
\author{ Wayne M. Saslow } 
\email{wsaslow@tamu.edu}
\affiliation{ Department of Physics, Texas A\&M University, College Station, TX 77843-4242 }
\date{\today}

\begin{abstract}
Many proposed and realized spintronic devices involve spin injection and accumulation at an interface between a ferromagnet and a non-magnetic material.  We examine the electric field, voltage profile, charge distribution, spin fluxes, and spin accumulation at such an interface.  We include the effects of both screening and spin scattering.  We also include both the spin-dependent chemical potentials ${\mu}_{\uparrow,\downarrow}$ and the effective magnetic field $\vec{H}^*$ that is zero in equilibrium.  For a Co/Cu interface, we find that the spin accumulation in the copper is an order of magnitude larger when both chemical potential and effective magnetic field are included.  We also show that screening contributes to the spin accumulation in the ferromagnet; this contribution can be significant. 
\end{abstract}



\maketitle

\section{Introduction}

Although electronic current has been studied since the early 19$^{\rm{th}}$ Century, spin current has been studied only much more recently.  In particular, spin transport across interfaces between metals and ferromagnets has been an important topic since the discovery of giant magnetoresistance (GMR),\cite{BaibichFertGMR,BinaschGrunbergGMR} 
the principle behind the predominant method of reading stored data. 
The magnetic read-head of a hard drive contains a thin non-magnetic layer sandwiched between two ferromagnetic layers. 

A theory for spin current and electrical potential at a metal/ferromagnet interface is given by Johnson and Silsbee\cite{JohnsonSilsbee} (JS); an appendix of that work is devoted to electric currents crossing such interfaces, and it considers the effect of the interfaces on spin fluxes and on electrical voltage.
Detailed theories for electrical currents crossing metal/ferromagnet multilayers (that is, series of interfaces) are given by Valet and Fert\cite{ValetFert} (VF), which includes solutions for the electric field and spin fluxes, and by Hershfield and Zhao\cite{Hershfield} (HZ).  
However, none of these theories considers semiconductors, and each makes a different assumption, not made by the present work, about some part of the \textit{magnetoelectrochemical} potential (first defined by JS and discussed in detail below).  
JS neglects the chemical potentials $\mu_{\uparrow} $ and $\mu_{\downarrow}$, HZ neglects the effective magnetic field\footnote{The scalar version of $\vec{H}^*$ is referred to by Ref.~\onlinecite{JohnsonSilsbee} as the \textit{magnetization potential}.} $\vec{H}^*$ (discussed below), and VF takes the chemical potential to be spin-independent.  

The present work revisits the problem of spin transport across the interface between a non-magnetic material (NM) and a ferromagnet (FM), and calculates the electric field, voltage, charge density, spin fluxes, and spin accumulation.  The results also apply to FM/FM and NM/NM interfaces.  We show that inclusion of both $\vec{H}^*$ and $\mu_{\uparrow} $ and $\mu_{\downarrow}$ are necessary to predict the spin accumulation near the interface.  For copper, neglecting either contribution decreases the spin accumulation by about a factor of ten.  Further, this work includes the surface screening mode (called the \textit{charge mode} by HZ), neglected by JS and VF, and ultimately neglected by HZ, which for large screening lengths (semiconductors) plays an important role in determining the spin current and the spin accumulation.  
Including the screening mode permits the electric field and potential to be continuous across the interface.  Previous works allow the field and potential to be discontinuous. 
Reference~\onlinecite{Fert01}, which extends VF by calculating the spin accumulation when the non-magnetic material is semiconducting, also neglects screening.\footnote{Reference~\onlinecite{Fert01} seems to take the chemical potential to be spin-independent, but this is not obvious.  It refers to the magnetoelectrochemical potential, $\bar{\mu}_{\uparrow,\downarrow} $, which is clearly spin-dependent, variously as the chemical potential (e.g., before its Eq.~(10)) and the electrochemical potential (e.g., before its Eq.~(1)), but it does not define it explicitly in terms of the ${\mu}_{\uparrow,\downarrow} $. However, because it defines the spin accumulation in a non-magnetic material (where $N_{\uparrow}=N_{\downarrow} $) to be proportional to the difference of its magnetoelectrochemical potentials (e.g., before its Eq.~(1), and in its Fig.~2), this agrees with the present Eq.~\eqref{mubarLin} if the chemical potential were spin-independent.}

Section~\ref{sec:TransEqs} briefly discusses the equations that govern spin-dependent transport in solids. 
Section~\ref{sec:Modes} finds the deviations from equilibrium due to the screening mode, the spin-diffusion mode, and a bulk response associated with the applied electric current.  
Section~\ref{sec:BCs} discusses the bulk and boundary conditions at an isolated interface. 
Section~\ref{sec:PreviousTheory} compares the assumptions of the present work to those of previous theories. 
Section~\ref{sec:IsoInt} gives the electric field, voltage, charge density, spin fluxes, and spin accumulation near a Co/Cu interface.  
Section~\ref{sec:Summary} provides a brief summary and conclusion.  
Appendix~\ref{appendix:SpinMode} shows detailed calculations for the spin-diffusion mode, the results of which are given in Sec.~\ref{sec:Modes}, and App.~\ref{appendix:IsolatedInt} explicitly gives the boundary conditions discussed in Sec.~\ref{sec:BCs}.

\section{Transport Equations}
\label{sec:TransEqs}

We use superscripts I and II or NM and FM to denote adjacent materials. 
When developing bulk equations that apply separately within each material, we omit the superscript, and reintroduce it when discussing materials in contact (or when discussing properties specific to a FM or a NM).

\subsection{Fundamental Relations}

Within each material, the number and current densities $n_{\uparrow,\downarrow}$ and $j_{\uparrow,\downarrow{_i}}$ are related by\cite{SearsSasHeating,Saslow2007}
\begin{align}
\partial_t n_{\uparrow} + \partial_i j_{\uparrow_{i}} = S, \quad \partial_t n_{\downarrow} + \partial_i j_{\downarrow_{i}} = -S. \label{Continuity}
\end{align}
Here $S$ is the rate at which down-spins flip to up-spins.  We consider the total electric current density $J=-ej_{\rm tot}$ to be a known uniform constant, and continuous across an interface.  For current along $x$ across an isolated interface (in the $yz$-plane) between materials I and II, we have $j_{\rm tot} =  j^{\rm (I)}_{\uparrow_{x}} + j^{\rm (I)}_{\downarrow_{x}} =  j^{\rm (II)}_{\uparrow_{x}} + j^{\rm (II)}_{\downarrow_{x}}$.

We take $\hat{M}$, the direction of the magnetization $\vec{M}$, to be fixed.  Since the electron g-factor is negative, for majority carriers defined to have up-spins, then $\hat{M}$ is aligned with the down-spins. 

We take $\bar{\mu}$ to be the {\it magnetoelectrochemical potential}, defined for up- and down-spin electrons as\cite{JohnsonSilsbee,Saslow2007}
\begin{align}
& \bar{\mu}_{\uparrow} = {\mu}_{\uparrow} - e \phi + \frac{|g| \mu_B}{2} \mu_0 \vec{H}^* \cdot \hat{M} ,\label{mubarup}\\
& \bar{\mu}_{\downarrow} = {\mu}_{\downarrow} - e \phi - \frac{|g| \mu_B}{2} \mu_0 \vec{H}^* \cdot \hat{M} , \label{mubardown}
\end{align}
where $\mu_{\uparrow}$ and $\mu_{\downarrow}$ are the respective chemical potentials of up- and down- spin carriers, $e>0$ is the magnitude of the electron charge, $\phi$ is the electrical potential, $g$ is the dimensionless g-factor (with $|g| \approx 2$ for electrons), $\mu_B$ is the Bohr magneton (with units of $J/T$), and $\mu_0$ is the permeability of free space\footnote{Reference~\onlinecite{JohnsonSilsbee} has a similar structure for magnetoelectrochemical potential, but its magnetic field term does not have $\mu_0$ due to its use of Gaussian units.} (with units of $N/A^2$).  
In the simplest case, $\vec{H}^*$ is the difference of the external field $\vec{H}_{0}$ and the uniform exchange field $\vec{H}_{int}$ (with $H_{int} \parallel \hat{M}$).  More generally, in addition to $\vec{H}_{0}$ we must include the magnetic dipole field $\vec{H}_{dip}$, the crystalline anisotropy field $\vec{H}_{an}$, and the non-uniform exchange field $\vec{H}_{ex}$ (proportional to $\nabla^2 \vec{M}$):\cite{SaslowSPIE}  
\begin{align}
\vec{H}^* = \vec{H}_{0} + \vec{H}_{dip}+\vec{H}_{an}+\vec{H}_{ex} - \vec{H}_{int}.
\label{Hstar}
\end{align}
We have $\vec{H}^*=\vec{0}$ in equilibrium.\footnote{As discussed in Ref.~\onlinecite{SaslowSPIE}, one can argue that $\vec{H}_{ex}$ is part of $\vec{H}_{int}$. This does not affect $\vec{H}^*$.}

By irreversible thermodynamics (see, for example, the general treatments in Refs.~\onlinecite{Onsager31}, \onlinecite{Prigogine}, and \onlinecite{deGrootTIP}, or the spin-related treatments of Refs.~\onlinecite{SearsSasHeating} and \onlinecite{Saslow2007}), the non-negativity of the rate of entropy production implies that the fluxes can be written in terms of thermodynamic forces, i.e., gradients of intensive thermodynamic quantities. Thus,
\begin{align}
j_{\uparrow_{i}} = -\frac{\sigma_{\uparrow}}{e^2} \partial_i \bar{\mu}_{\uparrow} - L_{\uparrow \downarrow} \partial_i \bar{\mu}_{\downarrow},\\
j_{\downarrow_{i}} =  -L_{\downarrow \uparrow} \partial_i \bar{\mu}_{\uparrow} - \frac{\sigma_{\downarrow}}{e^2} \partial_i \bar{\mu}_{\downarrow},
\end{align}
where $\sigma_{\uparrow}$ and $\sigma_{\downarrow}$ are the respective electrical conductivities of electrons of up- and down- spin, and the coefficients $L_{\downarrow \uparrow} = L_{\uparrow \downarrow}$ by the Onsager principle. 
We have implicitly neglected temperature gradients, which can also contribute to spin fluxes.\cite{SpinCal,GravierAnsermet06,HatamiBauer,SearsSasSSE} 
Neglecting the off-diagonal coefficients $L_{\downarrow \uparrow} = L_{\uparrow \downarrow}$, we have
\begin{align}
j_{\uparrow_{i}} = -\frac{\sigma_{\uparrow}}{e^2} \partial_i \bar{\mu}_{\uparrow}, \qquad j_{\downarrow_{i}} = - \frac{\sigma_{\downarrow}}{e^2} \partial_i \bar{\mu}_{\downarrow}. \label{Fluxes}
\end{align}

The non-negativity of the rate of entropy production gives\cite{Saslow2007,SearsSasHeating}
\begin{align}
S = -\alpha \left(\bar{\mu}_{\uparrow} - \bar{\mu}_{\downarrow} \right).
\label{FlipRate}
\end{align}
Here $\alpha \geq 0$ (with units of a density of states per second) is related to a characteristic spin-flip time (or, equivalently, to a characteristic spin-flip length).

We are interested in steady-state solutions, so that $\partial_t n_{\uparrow}=0=\partial_t n_{\downarrow}$. 
Taking the gradient of Eq.~\eqref{Fluxes} and employing Eqs.~\eqref{Continuity} and \eqref{FlipRate} then gives two coupled differential equations for $\bar{\mu}_{\uparrow}$ and $\bar{\mu}_{\downarrow}$,
\begin{gather}
-\frac{\sigma_{\uparrow}}{e^2} \partial^2_i \bar{\mu}_{\uparrow} = -\alpha \left(\bar{\mu}_{\uparrow} - \bar{\mu}_{\downarrow} \right), \label{muup}\\
 -\frac{\sigma_{\downarrow}}{e^2} \partial^2_i \bar{\mu}_{\downarrow} = \alpha \left(\bar{\mu}_{\uparrow} - \bar{\mu}_{\downarrow} \right). \label{mudown}
\end{gather}
On applying appropriate boundary conditions, Eqs.~\eqref{muup} and \eqref{mudown} give $\bar{\mu}_{\uparrow}$ and $\bar{\mu}_{\downarrow}$.  

\subsection{Linearized Relations} 
We are interested not only in $\bar{\mu}_{\uparrow}$ and $\bar{\mu}_{\downarrow}$, but also in $n_{\uparrow}$ and $n_{\downarrow}$ -- in particular, the difference of their deviations from equilibrium $\delta n_{\uparrow} - \delta n_{\downarrow}$, i.e., the \textit{spin accumulation} (which is proportional to the ``out-of-equilibrium magnetization'' or ``nonequilibrium magnetization'' discussed by VF and HZ).  Near equilibrium, we can linearize the deviations (denoted by $\delta$) from equilibrium of the chemical and magnetic contributions to the magnetoelectrochemical potentials: the chemical potential deviations can be written as
\begin{gather}
\delta \mu_{\uparrow} = \frac{\partial \mu_{\uparrow}}{\partial n_{\uparrow}} \delta n_{\uparrow},\quad \delta \mu_{\downarrow} = \frac{\partial \mu_{\downarrow}}{\partial n_{\downarrow}} \delta n_{\downarrow},
\end{gather}
and the deviation in the effective magnetic field at fixed $\vec{H}_{0}$ 
can be written as
\begin{align}
\delta \vec{H}^* \cdot \hat{M}=\frac{\mu_0 \delta \vec{M}}{\chi} \cdot \hat{M} = \frac{\mu_0 \mu_B}{\chi}(\delta n_{\uparrow} - \delta n_{\downarrow}),
\label{HstarTOn}
\end{align}
where $\chi$ is the magnetic susceptibility for an isotropic material (defined by $\chi_{ij} = \chi \delta_{ij}$).
Thus Eqs.~\eqref{mubarup} and \eqref{mubardown} give
 \begin{gather}
 \delta \bar{\mu}_{(\uparrow,\downarrow)} = \frac{\delta n_{(\uparrow,\downarrow)}}{N_{(\uparrow,\downarrow)}} - e \delta \phi \pm \frac{(\delta n_{\uparrow}-\delta n_{\downarrow})}{2 N_{\chi}},
 \label{mubarLin}
 \end{gather}
 where we define 
 \begin{align}
 N_{\uparrow} \equiv \frac{\partial n_{\uparrow}}{\partial \mu_{\uparrow}},\quad
 N_{\downarrow} \equiv \frac{\partial n_{\downarrow}}{\partial \mu_{\downarrow}} , \quad
 N_{\chi} \equiv \frac{\chi}{|g| \mu_B^2 \mu_0},
 \label{NupNdownZeta}
 \end{align}
 each of which has units of a density of states.

There are thus three unknowns ($\delta n_{\uparrow}$, $\delta n_{\downarrow}$, and $\delta \phi$). Eqs.~\eqref{muup} and \eqref{mudown} give two coupled differential equations, and Gauss's law provides a third:
\begin{align}
\partial_i^2  \delta \phi =& \frac{e}{\varepsilon_0 \varepsilon} \left(  \delta n_{\uparrow} +  \delta n_{\downarrow} \right).
\label{Gauss}
\end{align}
For the bulk response and each of the surface mode, we must find $\delta n_{\uparrow}$, $\delta n_{\downarrow}$, and $\delta \phi$.

\section{Static Bulk Response and Surface Modes}
\label{sec:Modes}
We now study the static bulk response and surface modes of the system.  
For brevity we write surface solutions to have the form $e^{-x/\ell}$ where $\ell$ is some length, although for the material on the left side of the interface one should use $e^{x}$ (because the deviations must decay as $x \rightarrow - \infty$).  
In general, each surface solution has the form $e^{\pm(x_{\rm int} -x)/\ell}$ where $x_{\rm int}$ is the position of the interface, but we take the interface to be at $x_{\rm int}=0$.


The electric field $\vec{E}$ and voltage $\phi$ are continuous everywhere.  (We call these ``Maxwell conditions.'')  To ensure this, we include the surface screening mode.  JS, VF, and HZ neglect screening and do not satisfy these conditions.   

We first discuss the bulk response associated with the electric current, which has a simpler structure than the surface modes associated with screening and with spin-diffusion.

\subsection{Bulk Respone ($dc$)}
We consider a system with a uniform constant electric current.  
The (bulk) response associated with this current, which can be thought of as a ``dc mode'' ($dc$), is characterized by a constant uniform electric field (which in principle differs for each material).  We define this field as
\begin{align}
\delta \vec{E}_{dc} \equiv E_{0_{dc}} \hat{x},
\label{Edc}
\end{align}
where $E_{0_{dc}}$ is a constant determined by applying boundary conditions.  
The potential associated with this mode is
\begin{align}
\delta \phi_{dc} = -E_{0_{dc}} x + V_{0_{dc}},
\label{phidc}
\end{align}
where $V_{0_{dc}}$ is another constant (with units of V) determined by applying boundary conditions.  By Gauss's Law there is no overall (bulk or surface) charge associated with this mode, as expected.  
Further, 
\begin{align}
\delta \bar{\mu}_{{\uparrow}_{dc}}=\delta \bar{\mu}_{{\downarrow}_{dc}}= -e \delta \phi_{dc} = eE_{0_{dc}}x-eV_{0_{dc}}.
\end{align}

Equation~\eqref{Fluxes} gives
\begin{align}
j_{{\uparrow}_{dc}} = -\frac{\sigma_{\uparrow}E_{0_{dc}}}{e}, \quad j_{{\downarrow}_{dc}} = -\frac{\sigma_{\downarrow}E_{0_{dc}}}{e}. \label{Fluxesdc}
\end{align}
Because $\sigma_{\uparrow}$ does not necessarily equal $\sigma_{\downarrow}$ (e.g., as for ferromagnets), there may be a non-zero spin current associated with the $dc$ mode.

\subsection{Screening Mode ($Q$)}
\label{subsec:Qmode}
One solution to Eqs.~\eqref{muup}, \eqref{mudown} and \eqref{Gauss} has $\delta \bar{\mu}_{\uparrow} = 0 = \delta \bar{\mu}_{\downarrow}$ so that $j_{\uparrow}=0=j_{\downarrow}$.  This mode is therefore entirely static (neither spin current nor charge current), corresponding to electric screening and characterized only by charge and potential gradients.  We therefore designate it the ``screening mode,'' and use the subscript $Q$ to denote its properties.\cite{*[{The screening mode is similarly discussed in }] [{, which considers surface screening by electrons and holes in a non-magnetic semiconductor.}] KrcmarSas}

Note that for metals the screening mode is not well-described by the present type of theory, but is instead associated with Friedel oscillations.\cite{LangKohn70,LangKohn71,Saslow72}  The following treatment of screening is more appropriate for doped semiconductors.

By Eq.~\eqref{mubarLin}, setting $\delta \bar{\mu}_{{\uparrow}_Q} - \delta \bar{\mu}_{{\downarrow}_Q}=0-0=0$ 
relates the up- and down-spin concentrations,
\begin{align}
N_{\downarrow} \left(N_{\uparrow} +  N_{\chi}\right) \delta n_{{\uparrow}_Q} = N_{\uparrow} \left(N_{\downarrow} +  N_{\chi} \right) \delta n_{{\downarrow}_Q},
\label{nupTOndownQ}
\end{align}
and setting $\delta \bar{\mu}_{{\uparrow}_Q} + \delta \bar{\mu}_{{\downarrow}_Q}=0+0=0$su gives
\begin{align}
\delta \phi_Q = \frac{1}{2e} \left( \frac{\delta n_{{\uparrow}_Q}}{N_{\uparrow} } +  \frac{\delta n_{{\downarrow}_Q}}{N_{\downarrow}} \right).
\label{phiTOnQ}
\end{align}

Define
\begin{gather}
N_{\alpha} \equiv N_{\uparrow} + N_{\downarrow} + 2N_{\chi},\label{Nalpha}
\end{gather}
with units of a density of states. 
Substitution of Eq.~\eqref{nupTOndownQ} into Eq.~\eqref{phiTOnQ} then yields
\begin{align}
N_{\uparrow} (N_{\downarrow} + N_{\chi}) \delta \phi_Q = \frac{N_{\alpha}}{2e} \delta n_{{\uparrow}_Q}.
\label{phiTOnupQ}
\end{align}
Substitution of Eqs.~\eqref{nupTOndownQ} and \eqref{phiTOnupQ} into Gauss's Law,  Eq.~\eqref{Gauss} then gives
\begin{align}
&N_{\alpha} \partial_i^2 \delta n_{{\uparrow}_Q}   =  \frac{2 e^2}{\varepsilon_0 \varepsilon} \left[  N_{\uparrow} \left( N_{\downarrow}+N_{\chi} \right) + N_{\downarrow} \left( N_{\uparrow} + N_{\chi} \right) \right] \delta n_{{\uparrow}_Q} .
\label{GaussQ1}
\end{align}
With the definitions
\begin{align}
&\ell_Q^2 \equiv  \frac{\varepsilon_0 \varepsilon}{2 e^2} \frac{N_{\alpha}}{N_{\beta}^2} ,
\label{lQ}\\
&N_{\beta}^2  \equiv 2 N_{\uparrow}N_{\downarrow} + N_{\chi} \left(N_{\uparrow} + N_{\downarrow} \right),  \label{Nbeta}
\end{align}
Eq.~\eqref{GaussQ1} can be written as
\begin{align}
&\partial_i^2 \delta n_{{\uparrow}_Q} = \frac{1}{\ell_Q^2} \delta n_{{\uparrow}_Q} .
\label{GaussQ2}
\end{align}

For $\chi \rightarrow \infty$ and $\varepsilon \rightarrow 1$, Eq.~\eqref{lQ} gives $\ell_Q^2 =  \varepsilon_0 /[e^2(N_{\uparrow} + N_{\downarrow})]$, which agrees with Ref.~\onlinecite{Hershfield}.

We now define the quantity $V_{0_Q}$ such that
\begin{align}
\delta n_{{\uparrow}_Q} \equiv 2e \frac{N_{\uparrow}}{N_{\alpha}} \left( N_{\downarrow} + N_{\chi} \right) V_{0_Q} e^{-x/\ell_Q},
\label{nupQ}
\end{align}
which satisfies Eq.~\eqref{GaussQ2}.
Then Eq.~\eqref{nupTOndownQ} gives
\begin{align}
&\delta n_{{\downarrow}_Q} = 2e \frac{N_{\downarrow}}{N_{\alpha}} \left( N_{\uparrow} + N_{\chi} \right) V_{0_Q} e^{- x/\ell_Q},\label{ndownQ}\\
&\delta \rho_Q = -e \left( \delta n_{{\uparrow}_Q} + \delta n_{{\downarrow}_Q}\right) \notag\\
&\quad \,\, \,\,= -2e^2 \frac{N_{\beta}^2}{N_{\alpha}} V_{0_Q} e^{- x/\ell_Q} = -\frac{\varepsilon_0 \varepsilon}{\ell_Q^2} V_{0_Q} e^{- x/\ell_Q}, \label{rhoQ}
\end{align}
and Eq.~\eqref{phiTOnQ} gives
\begin{align}
\delta \phi_Q =& V_{0_Q} e^{- x/\ell_Q}. \label{phiQ}
\end{align}

The screening mode can lead to a nonzero \textit{spin accumulation}, defined by 
\begin{gather}
\Delta n_{\sigma} \equiv \delta n_\uparrow - \delta n_\downarrow. 
\label{spinaccgen}
\end{gather}
Equations~\eqref{nupQ} and \eqref{ndownQ} give 
\begin{gather}
\Delta n_{\sigma_{Q}} 
= 2 \frac{N_{\chi}}{N_{\alpha}} (N_{\uparrow} - N_{\downarrow}) e V_{{0}_Q} e^{-x/\ell_Q},
\label{Qspinacc}
\end{gather}
which is nonzero in a ferromagnet, where $N_{\uparrow} \neq N_{\downarrow}$. 

\subsection{Spin Mode ($S$)}
\label{subsec:SMode}
The second solution to Eqs.~\eqref{muup}, \eqref{mudown} and \eqref{Gauss} is more complicated than the screening mode.  It is characterized by a nonzero spin current $j_\sigma \equiv j_{\uparrow} - j_{\downarrow}\neq 0$.  We therefore designate it the ``spin mode,'' and use the subscript $S$ to denote it.\cite{Hershfield}  Following Ref.~\onlinecite{Hershfield} we also use $Q$ (for charge) to denote the screening mode.  (The reader is thus warned that $S$ refers to spin, not to screening.)  

We now give the solution for the characteristic length, the spin concentrations, the electrical potential, and the spin accumulation associated with this mode.  The details of the analysis are given in Appendix~\ref{appendix:SpinMode}. 

Define the up- and down-spin associated lengths $\ell_{\uparrow_{S}}$ and $\ell_{\downarrow_{S}}$, which satisfy
 \begin{align}
 \ell_{{\uparrow}_S}^2 \equiv \frac{\sigma_{\uparrow}}{\alpha e^2},\qquad \ell_{{\downarrow}_S}^2 \equiv \frac{\sigma_{\downarrow}}{\alpha e^2}.
 \label{lsfupdown}
 \end{align}
 The decay length associated with the spin mode, variously called the ``spin-flip" or ``spin-diffusion" length, $\ell_{\rm sf}$, is then given by\cite{*[{The spin-diffusion length in a semiconductor may be more complicated when a field $E_{0}$ is applied; see }] [{.}] YuFlatte, *[{The spin-diffusion length is called $\delta_{i}$ by JS, $\ell_{\rm sf}$ by VF, and $\Lambda$ by }] [{; we follow VF by employing $\ell_{\rm sf}$.}] vSonvKempWyder} 
 \begin{align}
 \frac{1}{\ell_{\rm{{\rm{sf}}}}^2} =   \frac{1}{\ell_{{\uparrow}_S}^2} +   \frac{1}{\ell_{{\downarrow}_S}^2}. 
 \label{lsf} 
\end{align}
We also define
\begin{gather}
N_S \equiv \frac{\varepsilon_0 \varepsilon}{e^2\ell_{\rm{{\rm{sf}}}}^2} , \quad {C} \equiv  \frac{\ell_{\rm{{\rm{sf}}}}^2}{\ell_{{\uparrow}_S}^2} - \frac{\ell_{\rm{{\rm{sf}}}}^2}{\ell_{{\downarrow}_S}^2},
\label{NSC}
\end{gather}
where $N_{S}$ has units of a density of states and $C$ is dimensionless. 
With $V_{0_{S}}$ a constant to be determined by boundary conditions, the deviations in the electrical potential and up- and down- spin concentrations are then given by
\begin{align}
\delta \phi_S &= \left[ \frac{N_{\chi} \left(N_{\uparrow} - N_{\downarrow} \right) + {C} N_{\beta}^2  }
{N_S N_{\alpha} - 2 N_{\beta}^2} \right] V_{0_S} e^{-x/\ell_{\rm{sf}}} 
\notag\\&\equiv \xi V_{0_S} e^{-x/\ell_{\rm{sf}}}
,\label{phiS}\\
\delta n_{{\uparrow}_S} &= N_{\uparrow} e V_{0_S} e^{-x/\ell_{\rm{sf}}} \notag\\
& \times \left\{\frac{-2N_{\chi}N_{\downarrow} +N_S\left[ N_{\chi} + {C} \left( N_{\downarrow} +  N_{\chi} \right) \right] }
{ N_S N_{\alpha} - 2N_{\beta}^2  }\right\},\label{nupS}\\
\delta n_{{\downarrow}_S} &=  N_{\downarrow} e V_{0_S} e^{-x/\ell_{\rm{sf}}}\notag\\
& \times \left\{\frac{2N_{\chi}N_{\uparrow} +N_S\left[ -N_{\chi} + {C} \left( N_{\uparrow} +  N_{\chi} \right) \right] }
{ N_S N_{\alpha} - 2N_{\beta}^2  }\right\}.\label{ndownS}
\end{align} 
For a non-magnetic material, the dimensionless coefficient $\xi \rightarrow 0$.

The spin mode leads to a nonzero spin accumulation; Equations~\eqref{nupS} and \eqref{ndownS} give 
\begin{align}
&\Delta n_{\sigma{_S}} = e N_{\chi} V_{0_S} e^{-x/\ell_{\rm sf}}  \notag\\
&\times\left\{ \frac{-4 N_{\uparrow} N_{\downarrow} + N_S \left[N_{\uparrow} + N_{\downarrow} + {C} \left(N_{\uparrow} - N_{\downarrow} \right) \right]}
{ N_S N_{\alpha} - 2 N_{\beta}^2} \right\},
\label{SFMspinacc}
\end{align}
so that $\Delta n_{\sigma{_S}} $ is nonzero for both ferromagnets and non-magnetic materials.  For the latter, Eq.~\eqref{SFMspinacc} simplifies to
\begin{align}
\Delta n_{\sigma_{S}}^{\rm (NM)} =\frac{N_{\uparrow} N_{\chi} e V_{0_S}}{\displaystyle N_{\uparrow}+ N_{\chi} } e^{-x/\ell_{\rm{{\rm{sf}}}}}.
\label{SNMspinacc}
\end{align}

The spin-carrier currents associated with the spin mode are given by 
\begin{align}
&\delta j_{\uparrow{_S}} = -\delta j_{\downarrow{_S}} = \left(\frac{\sigma_{\uparrow} \sigma_{\downarrow}}{\sigma_{\uparrow}  + \sigma_{\downarrow}} \right) \frac{V_{0{_S}}}{e \ell_{\rm sf}} e^{-x/\ell_{sf}}. \label{FluxesS} 
\end{align}
The total electric current $-e \delta j_{\rm tot} = -e (\delta j_{\uparrow} + \delta j_{\downarrow})=0$ for the spin mode, but there is a nonzero spin current $ \delta j_{\sigma} \equiv \delta j_{\uparrow} - \delta j_{\downarrow}$, given by
\begin{align}
&\delta j_{\sigma{_S}}  =2 \left(\frac{\sigma_{\uparrow} \sigma_{\downarrow}}{\sigma_{\uparrow}  + \sigma_{\downarrow}} \right) \frac{V_{0{_S}}}{e \ell_{\rm sf}} e^{-x/\ell_{sf}}. \label{SpinFluxS}
\end{align}

\subsection{Description Near Interface}

A full description of the region near an interface involves the combination of both surface modes (S and Q) derived above, and the bulk constant current (dc) mode.  For the potential, electric field, charge density near an interface located at $x=x_{\rm{int}}$, from Eqs.~\eqref{phiS}, \eqref{rhoS}, \eqref{rhoQ}, \eqref{phiQ}, \eqref{Edc}, and \eqref{phidc} we have, with four unknowns per material ($E_{0_{dc}}$, $V_{0_{dc}}$, $V_{0_Q}$, and $V_{0_S}$) to be determined by boundary conditions,
\begin{align}
&\delta \phi = \xi V_{0_S} e^{\pm (x-x_{\rm{int}})/\ell_{\rm{sf}}} + V_{0_Q} e^{\pm (x-x_{\rm{int}})/\ell_{Q}} \notag\\
&\qquad  - E_{0_{dc}}(x-x_{\rm{int}}) +V_{0_{dc}},\label{PhiTot}\\
&\delta E = \mp \frac{\xi V_{0_S}}{\ell_{\rm{sf}}} e^{\pm (x-x_{\rm{int}})/\ell_{\rm{sf}}} \mp \frac{V_{0_Q}}{\ell_{Q}} e^{\pm (x-x_{\rm{int}})/\ell_{Q}} + E_{0_{dc}},\label{ETot}\\
&\delta \rho = - \epsilon_0 \epsilon \left( \frac{\xi V_{0_S}}{\ell_{\rm{sf}}^2}e^{\pm (x-x_{\rm{int}})/\ell_{\rm{sf}}} + \frac{V_{0_Q}}{\ell_{Q}^2}e^{\pm (x-x_{\rm{int}})/\ell_{Q}} \right). \label{rhoTot}
\end{align}
The top (bottom) sign corresponds to the material on the left (right) of the interface.  

The contributions to the total electric current from the surface modes is zero, as expected, so that Eq.~\eqref{Fluxesdc} gives the electric current to be everywhere given by
\begin{gather}
J = -e j_{\rm tot} = - \left(\sigma_{\uparrow} + \sigma_{\downarrow}\right) E_{0_{dc}}.
\label{JTot}
\end{gather}
The spin mode does contribute to the nonconserved spin-up, spin-down, and total spin currents, which, combining Eqs.~\eqref{FluxesS} and \eqref{Fluxesdc}, are given by
\begin{align}
& j_{\uparrow} = - \frac{\sigma_{\uparrow}  E_{0_{dc}}}{e} + \left(\frac{\sigma_{\uparrow} \sigma_{\downarrow}}{\sigma_{\uparrow} + \sigma_{\downarrow}} \right) \frac{V_{0{_S}}}{e \ell_{\rm sf}} e^{\pm(x-x_{\rm int})/\ell_{\rm sf}}
\label{jupTot},\\
& j_{\downarrow} = - \frac{\sigma_{\downarrow} }{e} E_{0_{dc}} - \left(\frac{\sigma_{\uparrow} \sigma_{\downarrow}}{\sigma_{\uparrow} + \sigma_{\downarrow}} \right) \frac{V_{0{_S}}}{e \ell_{\rm sf}} e^{\pm(x-x_{\rm int})/\ell_{\rm sf}}
\label{jdownTot},\\
&j_{\sigma} = - \left( \frac{\sigma_{\uparrow} - \sigma_{\downarrow}}{e} \right) E_{0_{dc}} + \left(\frac{2 \sigma_{\uparrow} \sigma_{\downarrow}}{\sigma_{\uparrow} + \sigma_{\downarrow}} \right) \frac{V_{0{_S}}}{e \ell_{\rm sf}} e^{\pm(x-x_{\rm int})/\ell_{\rm sf}}
\label{jspinTot}.
\end{align}
There is no contribution from $V_{0{_Q}}$ because there are no carrier currents associated with the charge mode.

For the spin accumulation, Eqs.~\eqref{SNMspinacc}, \eqref{SFMspinacc}, and \eqref{Qspinacc} yield 
\begin{align}
\Delta n_{\sigma} =& e N_{\chi} \xi V_{0_S} e^{\pm (x-x_{\rm{int}})/\ell_{\rm sf}} \notag\\
\times &\left\{ \frac{-4 N_{\uparrow} N_{\downarrow} + N_S \left[N_{\uparrow} + N_{\downarrow} + R \left(N_{\uparrow} - N_{\downarrow} \right) \right]}{ N_\chi \left( N_{\uparrow} -N_{\downarrow}\right) + R N_{\beta}^2} \right\} \notag\\
&+ 2e \frac{N_{\chi}}{N_{\alpha}} (N_{\uparrow} - N_{\downarrow}) V_{{0}_Q} e^{\pm (x-x_{\rm{int}})/\ell_Q} \label{FMspinaccTot}.
\end{align}
For a non-magnetic material this simplifies to
\begin{align}
\Delta n_{\sigma}^{\rm (NM)} =& \frac{N_{\uparrow} N_{\chi} e V_{0_S}}{\displaystyle N_{\uparrow}+ N_{\chi} } e^{\pm (x-x_{\rm{int}})/\ell_{\rm{{\rm{sf}}}}} 
\label{NMspinaccTot}.
\end{align}

\section{Boundary and Bulk Conditions}
\label{sec:BCs}

\begin{figure}[t]
\centerline{\includegraphics[width=0.60\textwidth]{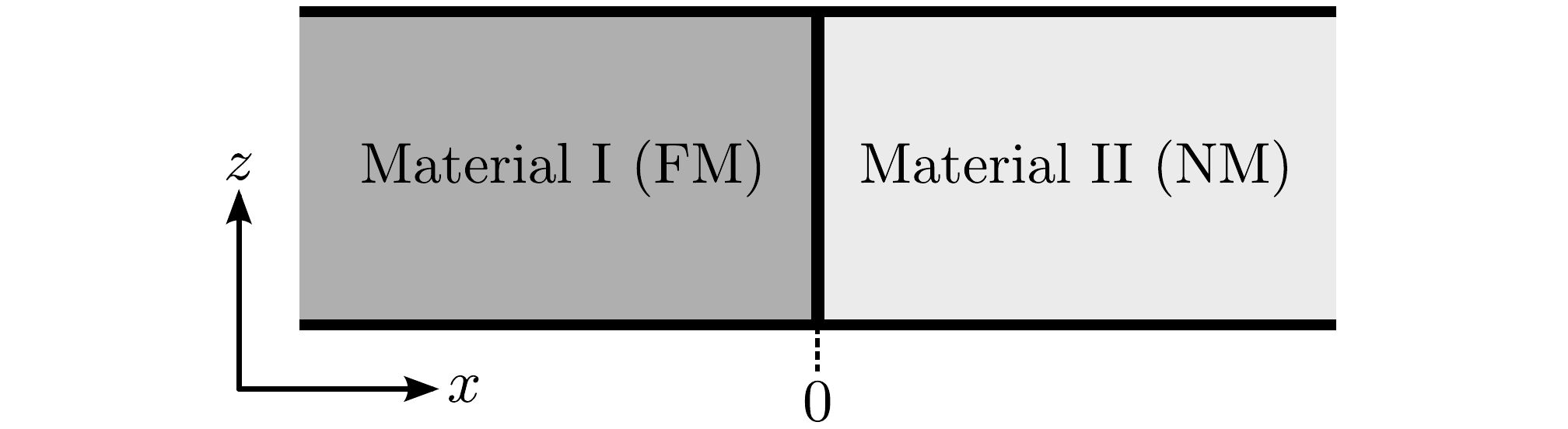}}
\caption[An isolated interface between a ferromagnet and a non-magnetic material.]
{An isolated interface between a ferromagnet (dark gray, at $x<0$) and a non-magnetic material (light gray, at $x>0$).  This work considers an electric current density $J\hat{x}$, and magnetization of the FM along $\pm \hat{z}$.  }
\label{fig:Geometry} 
\end{figure} 

For an isolated interface at $x_{\rm{int}}=0$ (see Fig.~\ref{fig:Geometry}) between materials I (at $x<0$) and II (at $x>0$), in general there are eight unknowns ($E_{0_{dc}}$, $V_{0_{dc}}$, $V_{0_Q}$, and $V_{0_S}$ for each of materials I and II).  
There are eight conditions:\footnote{Reference~\onlinecite{Hershfield} similarly numbers and discusses the conditions necessary to solve for the unknowns at such a boundary.  It uses conditions {\bf (iii-vii)} of the present work as its conditions (1-5), although in a different order. Furthermore, it makes use of the present work's condition {\bf (viii)}, though it does not number it.  However, because it neglects the screening mode at the interface, it does not apply the Maxwell conditions {\bf (i-ii)}.} 
\begin{description}
\item[(i-ii)] the potential $\phi$ and field $\vec{E}$ are continuous across the interface -- Maxwell conditions; 
\item[(iii)] the electric current $-e(j_{\uparrow} + j_{\downarrow})$ is continuous across the interface -- charge conservation; 
\item[(iv)] the spin current is assumed continuous across the interface (although we take both up- and down-spin currents to be continuous, this is only a single condition since condition {\bf (iii)} constrains their sum) -- assumption of no surface spin-scattering; 
\item[(v-vi)] the up- and down-spin currents across the interface are directly proportional to the discontinuity in up- and down-spin magnetoelectrochemical potential across the interface\cite{JohnsonSilsbee,SearsSasHeating,Saslow2007,*[{A theory for the polarization of spin-currents injected from a FM into a 2DEG is given by }] [{. This work takes zero surface resistance.}] Schmidt00}  -- irreversible thermodynamics; 
\item[(vii)] the total electric current $-e(j_{\uparrow} + j_{\downarrow})$ is a known constant; 
and 
\item[(viii)] there is an arbitrary constant voltage (which we define by setting the voltage $V_{0_{dc}}^{\rm{(II)}} \equiv 0$).
\end{description}
Conditions {\bf(i)}-{\bf(vi)} are boundary conditions and {\bf(vii)} and {\bf(viii)} are bulk conditions.  They are explicitly calculated in Appendix~\ref{appendix:IsolatedInt}. 

For a multilayer (a series of $k$ interfaces between $k+1$ materials), each additional interface adds another of each of the boundary conditions {\bf(i)}-{\bf(vi)}, so that in general there are $6k + 2$ conditions.

\section{Comparison to Previous Theories}
\label{sec:PreviousTheory}

As noted above, the theories of Johnson and Silsbee (JS), Valet and Fert (VF) and Hershfield and Zhao (HZ) neglect the screening mode, and therefore cannot have field and potential continuity at the interface.  Further, JS neglects the chemical potentials $\mu_{\uparrow}$ and $\mu_{\downarrow}$ and HZ neglects the internal magnetic field $\vec{H}^*$.  
The discrepancy between predicted spin accumulation, found below, particularly in a non-magnetic material, demonstrates that inclusion of all parts of the magnetoelectrochemical potential is essential for calculating the spin accumulation in a non-magnetic material, even to within an order of magnitude.  For comparison of Eq.~\eqref{SNMspinacc} to the spin accumulation predicted for these other works $\rm W=HZ$ and $\rm W=HZ$, we define the dimensionless factor $\zeta_{{}_{\rm W}}$ as
 \begin{gather}
 \left[\Delta n_{\sigma_{S}}^{\rm (NM)}\right]_{{}_{\rm W}}  = \zeta_{{}_{\rm W}} \Delta n_{\sigma_{S}}^{\rm (NM)}.
 \label{SpinAccCompare}
 \end{gather}
Note that $\zeta_{{}_{\rm W}}=1$ for the present work.  We show below that if one of $\zeta_{{}_{\rm HZ}}$ or $\zeta_{{}_{\rm JS}}$ is near unity (and therefore agrees with the present work), then  the other diverges, so that at least one of the assumptions gives results that significantly disagree with the present work.

\subsection{Neglecting $\vec{H}^*$ and the Screening Mode}
Neglecting the last term (proportional to $ \vec{H}^* \cdot \hat{M}$) in Eq.~\eqref{mubarLin}, as in HZ,\cite{Hershfield} is equivalent to taking $\chi \rightarrow \infty$ (and therefore $N_\chi \rightarrow \infty$) in the present results.  Under this assumption, $N_\alpha \rightarrow 2N_{\chi}$ and $N_{\beta}^2 \rightarrow N_{\chi} (N_{\uparrow} + N_{\downarrow})$. 
Equations~\eqref{phiS}-\eqref{ndownS} then simplify to 
\begin{align}
&\delta \phi_{S}^{\rm HZ} = \left[\frac{N_{\uparrow} - N_{\downarrow} + {C} (N_{\uparrow} + N_{\downarrow})}{2 \left(N_{S}-N_{\uparrow}-N_{\downarrow} \right)} \right]V_{0_S} e^{-x/\ell_{sf}},\label{phiSHZ}\\
&\delta n_{{\uparrow}_{S}}^{\rm HZ} = N_{\uparrow} e V_{0_S} e^{-x/\ell_{sf}} \left[ \frac{-2 N_{\downarrow} + N_S(1 + {C})}{2 \left(N_{S}-N_{\uparrow}-N_{\downarrow} \right)}\right],\label{nupSHZ}\\
&\delta n_{{\downarrow}_{S}}^{\rm HZ} = N_{\downarrow} e  V_{0_S} e^{-x/\ell_{sf}} \left[ \frac{2 N_{\uparrow} + N_S (-1 + {C})}{2 \left(N_{S}-N_{\uparrow}-N_{\downarrow} \right)} \right]. \label{ndownSHZ}
\end{align}
HZ neglect the screening mode, so the spin-diffusion mode is the only surface mode, and it gives a spin accumulation of
\begin{align}
&\Delta n_{\sigma}^{\rm HZ} =e V_{0{_S}} e^{-x/\ell_{sf}} \notag\\
&\times \left\{  \frac{-4 N_{\uparrow}N_{\downarrow} + N_{S} \left[ N_{\uparrow} + N_{\downarrow} + {C} \left( N_{\uparrow} - N_{\downarrow} \right) \right] }{2 \left(N_{S}-N_{\uparrow}-N_{\downarrow} \right)}  \right\}. \label{SpinAccHZ}
\end{align}

\begin{table}[t]
\caption[Bulk and interfacial properties of cobalt and copper, and well-known constants.]{Bulk and interfacial properties of cobalt and copper, and well-known constants.  Here, $A$ is the area of the interface, and $R$ is the spin-dependent interface resistance.  $^\dagger$Value is for the (100) orientation.  $^\ddagger$The susceptibility of Cobalt is field-dependent, with $70 \leq \chi^{\rm Co} \leq 250$ (see Table~2.2 of Ref.~\onlinecite{EMShielding}); we take an intermediate value.}
\label{tab:Numeric1}
\begin{tabularx}{0.46\textwidth}{l r X l c}
\specialrule{1pt}{1pt}{1pt}
\specialrule{1pt}{0.5pt}{2pt}
Quantity & Value & & Units & Ref\\ 
\specialrule{0.8pt}{2pt}{4pt}
$\sigma_{\uparrow}^{{\rm Co}}$ & \hspace*{6pt} $\displaystyle 2.47 \times 10^{7}$ & \hspace*{6pt} &   $\Omega^{-1}$-m$^{-1}$    \hspace*{6pt}& \cite{StilesXiaoZangwill04}\\ [1ex]
$\sigma_{\downarrow}^{{\rm Co}}$ & $\displaystyle 0.913 \times 10^{7}$ &  &  $\Omega^{-1}$-m$^{-1}$ & \cite{StilesXiaoZangwill04}\\ [1ex]
$\sigma_{\downarrow}^{{\rm Cu}}$, $\sigma_{\uparrow}^{{\rm Cu}}$ & $\displaystyle 8.35 \times 10^{7}$ &  &  $\Omega^{-1}$-m$^{-1}$ & \cite{StilesXiaoZangwill04}\\[1ex]
${\ell_{\rm{sf}}^{{\rm Co}}}$ & $59 \times 10^{-9}$ & & m & \cite{StilesXiaoZangwill04}\\ [1ex]
${\ell_{\rm{sf}}^{{\rm Cu}}}$ & $450 \times 10^{-9}$ & & m & \cite{StilesXiaoZangwill04}\\[1ex]
$N_{\uparrow}^{{\rm Co}}$  & $\displaystyle 5.10 \times 10^{46}$ & &  J$^{-1}$--m$^{-3}$ & \cite{StilesXiaoZangwill04}\\ [1ex]
$N_{\downarrow}^{{\rm Co}}$  & $\displaystyle 19.7 \times 10^{46}$ & &  J$^{-1}$--m$^{-3}$ & \cite{StilesXiaoZangwill04}\\[1ex]
$N_{\uparrow}^{{\rm Cu}}$, $N_{\downarrow}^{{\rm Cu}}$  \hspace*{18pt} & $\displaystyle 3.89 \times 10^{46}$ & &  J$^{-1}$--m$^{-3}$ & \cite{StilesXiaoZangwill04}\\ [1ex]
$A R^{{\rm Cu/Co}}_{\uparrow}$ & $0.31 \times 10^{-15}$ &  &  $\Omega$--m$^2$&  \cite{StilesPenn}$^\dagger$ \\ [1ex]
$A R^{{\rm Cu/Co}}_{\downarrow}$ & $2.31 \times 10^{-15}$ & &  $\Omega$--m$^2$& \cite{StilesPenn}$^\dagger$ \\[1ex]
$\chi^{{\rm Co}}$ & $\approx 100$ &  & &  \cite{EMShielding}$^{\ddagger}$ \\   [1ex]
$\chi^{{\rm Cu}}$ & $-0.932 \times 10^{-5}$ &  & & \cite{Bowers} \\  [1ex]
$\mu_B$ & \hspace*{12pt} $9.27 \times 10^{-24}$  &  & J--T$^{-1}$ & \\ [1ex]
$\mu_0$ & $4 \pi  \times 10^{-7}$ & & N -- A$^{-2}$ & \\  [1ex]
$\varepsilon_0$ & $8.85  \times 10^{-12}$ & & A--s--V$^{-1}$--m$^{-1}$ & \\  [1ex]
$e$ & $1.6 \times 10^{-19}$ & & C & \\    [1ex] 
$|g|$ & $\approx 2$ & & & \\    
\specialrule{1pt}{1pt}{0.5pt}
\specialrule{1pt}{1pt}{2pt}
\end{tabularx}
\end{table}
 
\begin{table}[ht]
\caption[Bulk and interfacial properties of cobalt and copper, calculated from the results of the present work.]
{Bulk and interfacial properties of cobalt and copper, calculated from the results of the present work (and Table~\ref{tab:Numeric1}).  Here $\alpha$ is found from Eq.~\eqref{alpha}.  }
\begin{tabularx}{0.46\textwidth}{l r X l}
\specialrule{1pt}{1pt}{1pt}
\specialrule{1pt}{0.5pt}{2pt}
Quantity & Value & & Units\\ 
\specialrule{0.8pt}{2pt}{4pt}
$\sigma^{\rm Cu}\equiv \sigma_{\uparrow}^{\rm Cu}+\sigma_{\downarrow}^{\rm Cu}$ \hspace*{12pt} & $16.7 \times 10^{7}$ &  & $\Omega^{-1}$-m$^{-1}$ \\  [1ex]
$g_{\uparrow}$ & $3.23 \times 10^{15}$ & & $\Omega^{-1}$-m$^{-2}$ \\  [1ex]
$g_{\downarrow}$ & $0.433 \times 10^{15}$ & & $\Omega^{-1}$-m$^{-2}$ \\ [1ex]
$N_\chi^{{\rm Co}}$ & $4.63 \times 10^{53}$ & & J$^{-1}$--m$^{-3}$ \\ [1ex]
$N_\chi^{{\rm Cu}}$ & $-4.32 \times 10^{46}$ & & J$^{-1}$--m$^{-3}$\\  [1ex]
$N_S^{{\rm Co}}$ & $9.93 \times 10^{40}$ & & J$^{-1}$--m$^{-3}$\\ [1ex]
$N_S^{{\rm Cu}}$ & $1.71 \times 10^{39}$ & & J$^{-1}$--m$^{-3}$\\ [1ex]
$N_\alpha^{{\rm Co}}$ & $9.26 \times 10^{53}$ & & J$^{-1}$--m$^{-3}$\\  [1ex]
$N_\alpha^{{\rm Cu}}$ & $-0.851 \times 10^{46}$ & & J$^{-1}$--m$^{-3}$\\ [1ex]
${N_\beta^{{\rm Co}}}^2$ & $1.15 \times 10^{101}$ & & J$^{-2}$--m$^{-6}$\\ [1ex]
${N_\beta^{{\rm Cu}}}^2$ & $-3.31 \times 10^{92}$ & & J$^{-2}$--m$^{-6}$\\ [1ex]
$\alpha^{{\rm Co}}$ & $74.8 \times 10^{57}$ & & J$^{-1}$--m$^{-3}$--s$^{-1}$\\  [1ex]
$\alpha^{{\rm Cu}}$ & $8.05 \times 10^{57}$ & & J$^{-1}$--m$^{-3}$--s$^{-1}$\\  [1ex]
$\ell_{{\uparrow}_{S}}^{\rm Co}$ & $114 \times 10^{-9}$ & & m\\ [1ex]
$\ell_{{\downarrow}_{S}}^{\rm Co}$ & $69.0 \times 10^{-9}$ & & m\\ [1ex]
$\ell_{{\uparrow}_{S}}^{\rm Cu}, \ell_{{\downarrow}_{S}}^{\rm Cu}$ & $636 \times 10^{-9}$ & & m \\ [1ex]
${\ell_Q^{{\rm Co}}}$ & $0.0373 \times 10^{-9}$ & & m\\ [1ex]
${\ell_Q^{{\rm Cu}}}$ & $0.0667 \times 10^{-9}$ & & m\\  [1ex]
${C}^{\rm Co}$ & $-0.460$ & & \\ [1ex]
${C}^{\rm Cu}$ & $0$ & & \\ [1ex]
$\xi^{\rm Co}$ & $0.524$ & & \\[1ex]
$\xi^{\rm Cu}$  & $0$ & & \\
\specialrule{1pt}{1pt}{0.5pt}
\specialrule{1pt}{1pt}{2pt}
\end{tabularx}
\label{tab:Numeric2}
\end{table}

Direct comparison can be made to the results of the present work in the non-magnetic material.  With $\zeta_{{}_{\rm W}} $ defined by Eq.~\eqref{SpinAccCompare}, we have
\begin{gather}
\zeta_{{}_{\rm HZ}} = (1 + N_{\uparrow}/N_{\chi}).
\label{zetaHZ}
\end{gather}
Using Tables~\ref{tab:Numeric1} and \ref{tab:Numeric2}, we find $\zeta_{{}_{\rm HZ}} \approx 0.0986$ for Cu. (Cu is a diamagnet, therefore it has $N_{\chi}<0$; for a paramagnet, where $N_{\chi}>0$, the underestimation of spin accumulation for the HZ assumptions is less striking, although it remains significant.)  For the ferromagnet, the spin accumulation due to the screening mode is neglected, and the spin accumulation due to the spin mode agrees with the present work to within the precision of the present calculations.   Hence, the assumptions made by HZ seem appropriate for ferromagnets but not for nonmagnetic materials.

\subsection{Neglecting $\mu_{\uparrow}$, $\mu_{\downarrow}$ and the Screening Mode}
JS neglects the chemical potentials $\mu_{\uparrow}$ and $\mu_{\downarrow}$ in Eq.~\eqref{mubarLin}, which is equivalent to taking $N_{\uparrow,\downarrow} \rightarrow \infty$ in the present work. 
It also neglects the screening mode. Various properties of the spin mode are now calculated under these assumptions. 

Equation~\eqref{phiS} gives
\begin{equation}
\delta \phi_{S}^{\rm JS} = -\frac{{C}}{2}V_{0_S}e^{-x/\ell_{sf}}.
\label{phiSJS}
\end{equation}
Further, Eqs.~\eqref{nupS} and \eqref{ndownS} give
\begin{align}
&\delta n_{\uparrow{_S}}^{\rm JS} = \left( \frac{N_{\chi}}{2} - \frac{{C} N_{S}}{4} \right) e V_{0{_S}} e^{-x/\ell_{sf}} ,\label{nupSJS}\\
&\delta n_{\uparrow{_S}}^{\rm JS} = -\left( \frac{N_{\chi}}{2} + \frac{{C} N_{S}}{4} \right) e V_{0{_S}} e^{-x/\ell_{sf}} ,\label{ndownSJS}
\end{align}
so that the spin accumulation is given by
\begin{align}
\Delta n_{\sigma}^{\rm JS} = N_{\chi} e V_{0_S} e^{-x/\ell_{sf}} .
\label{SpinAccVF}
\end{align}

Direct comparison can be made to the results of the present work in the non-magnetic material.  With $\zeta_{{}_{\rm W}} $ defined by Eq.~\eqref{SpinAccCompare}, we have
\begin{gather}
\zeta_{{}_{\rm JS}} = (1 + N_{\chi}/N_{\uparrow}).
\label{zetaJS}
\end{gather}
Using Tables~\ref{tab:Numeric1} and \ref{tab:Numeric2}, we find $\zeta_{{}_{\rm JS}} \approx -0.109$ for Cu.  Thus, the JS assumptions seem inappropriate for determining the spin accumulation in non-magnetic materials, particularly those that are diamagnetic.  

Note that Eqs.~\eqref{zetaHZ} and \eqref{zetaJS} preclude simultaneously having $\zeta_{{}_{\rm JS}} \approx 1$ and $\zeta_{{}_{\rm HZ}} \approx 1$.

\section{Co/Cu Interface}
\label{sec:IsoInt}

\begin{figure}[t]
\centerline{\includegraphics[width=0.46\textwidth]{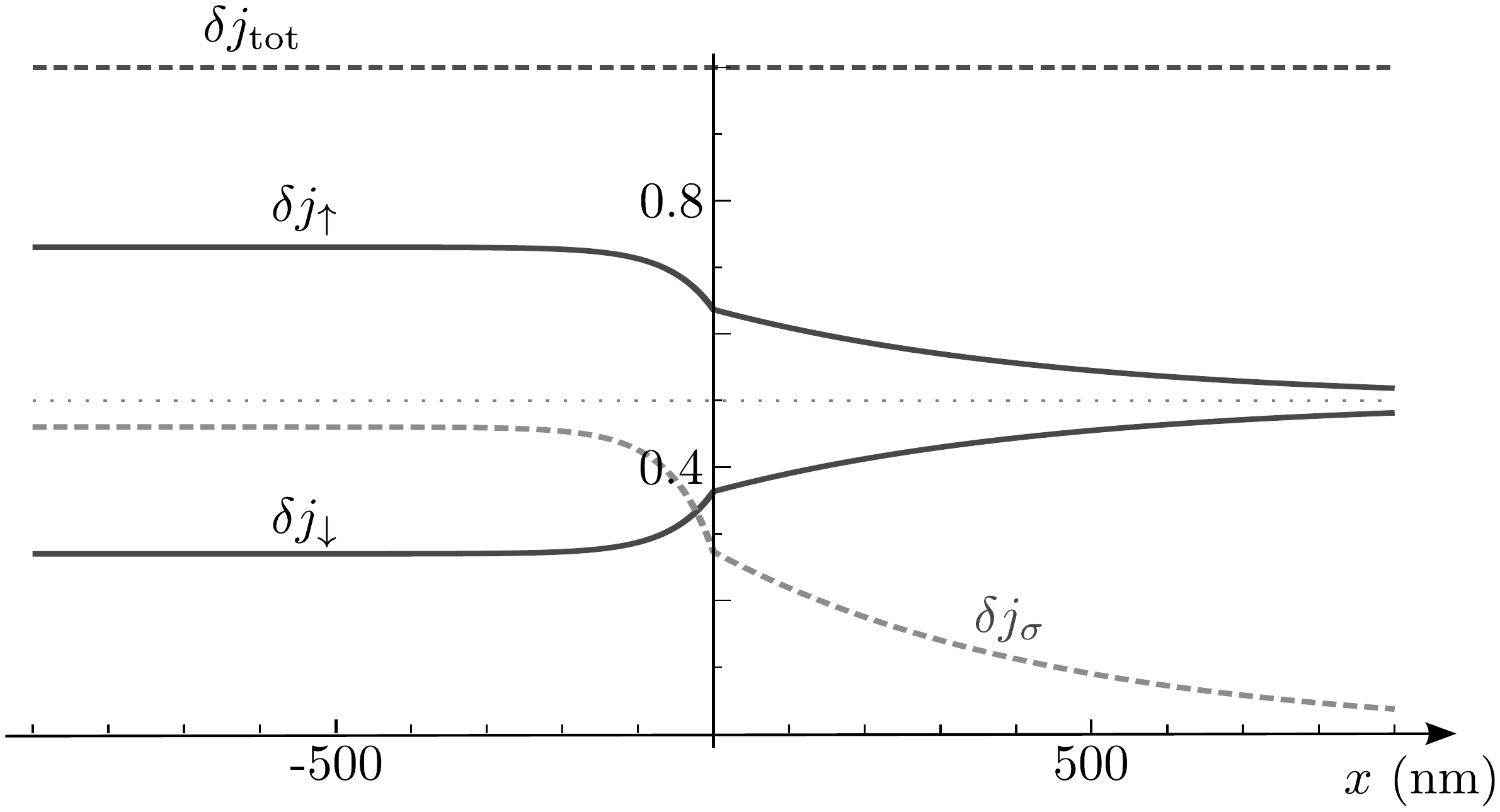}}
\caption
{The spin-up and spin-down carrier fluxes $\delta j_{\uparrow}$ and $\delta j_{\downarrow}$, and the total flux $\delta j_{\rm tot}$ and spin flux $\delta j_{\sigma}$, given by Eqs.~\eqref{JTot}-\eqref{jspinTot}, with the (uniform) total flux normalized to unity, near an interface between cobalt ($x<0$) and copper ($x>0$) vs. $x$ (in nm).  The grey dotted line marks 0.5, that is, half of the total current. 
}
\label{fig:IsoIntCurrent} 
\end{figure}

For an isolated interface (as in Fig.~\ref{fig:Geometry}), Appendix~\ref{appendix:IsolatedInt} uses each of the above conditions to find an explicit equation for the eight unknowns and writes the unknowns in terms of dimensionless variables. 
We now present numerical results for the spin fluxes (see Fig.~\ref{fig:IsoIntCurrent}), voltage, electric field, charge density, and spin accumulation, for a cobalt/copper interface, with material parameters given by Tables~\ref{tab:Numeric1} and \ref{tab:Numeric2}.

\begin{figure}[t]
\centerline{\includegraphics[width=0.46\textwidth]{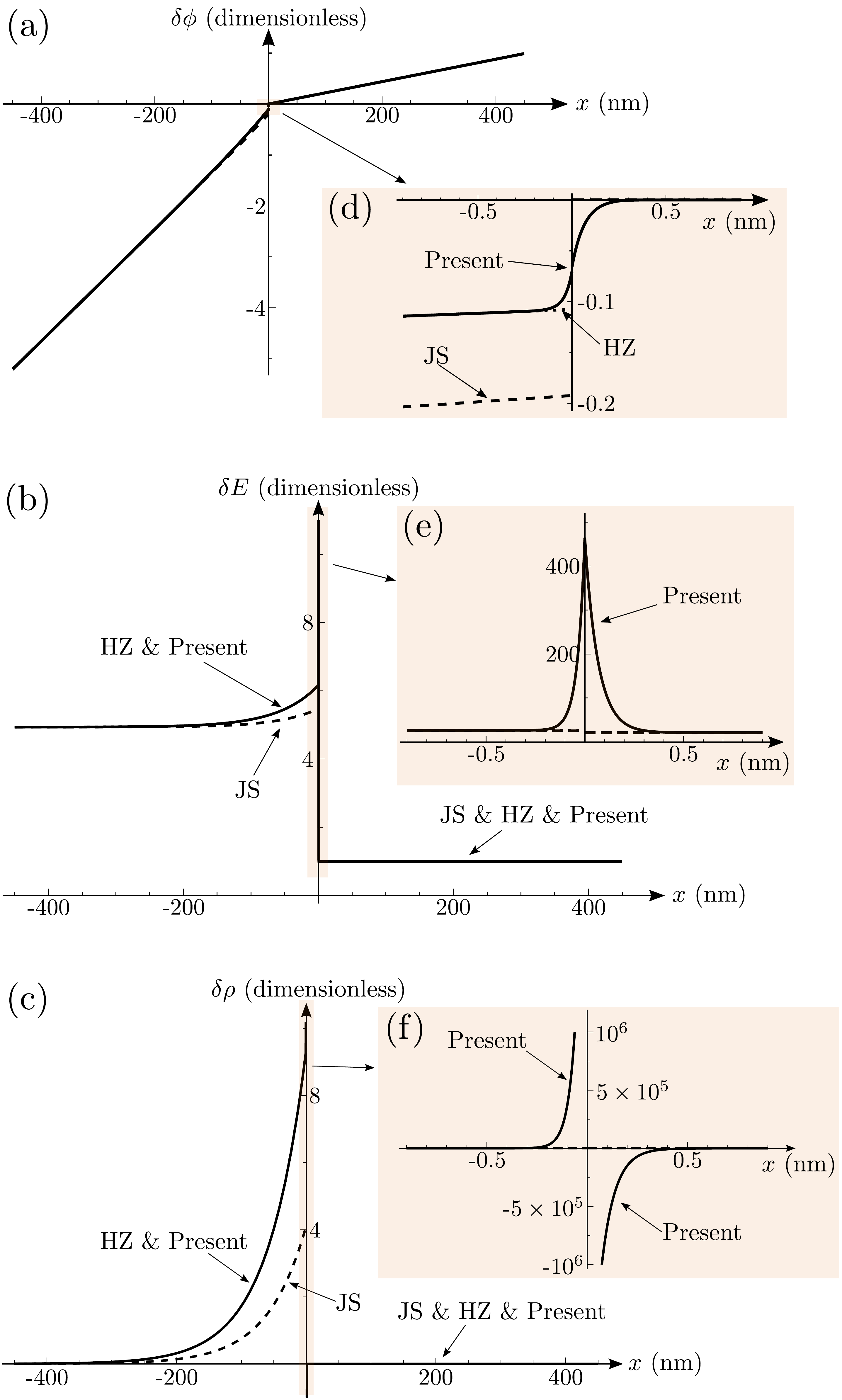}}
\caption
{The dimensionless electrical potential, field, and charge density (given by Eqs.~\eqref{PhiTot}-\eqref{rhoTot}) in arbitrary units near an interface between cobalt ($x<0$) and copper ($x>0$) vs. $x$ (in nm).  
Within 1~$\mu$m (a,b,c), and within 1~nm (d,e,f), of the interface. 
Solid line -- present work; dashed line -- JS; dotted line -- HZ.  
HZ coincides closely with the present work except within a screening length of the interface. 
JS gives somewhat different results in the cobalt within a spin-diffusion length of the interface.
}
\label{fig:IsoIntElectrical} 
\end{figure} 

%
%
%
%
%

Figures~\ref{fig:IsoIntElectrical}a-\ref{fig:IsoIntElectrical}c show that, outside of a screening length $\ell_{Q}$ of the interface, the electrical potential, field and charge nearly coincide for the present work and HZ, with JS showing discrepancies near the interface in the ferromagnet ($x<0$).   However, the present work significantly differs from JS and HZ within a screening length of the interface, as seen in  Figs.~\ref{fig:IsoIntElectrical}d-\ref{fig:IsoIntElectrical}f.  
Figures~\ref{fig:IsoIntElectrical}d and \ref{fig:IsoIntElectrical}e show, for the present work, the continuity of the electrical potential and field at the interface.  They also show, for HZ and JS, the discontinuities in the potential and field (due to scale, these field discontinuities are more obvious in Fig.~\ref{fig:IsoIntElectrical}b than in Fig.~\ref{fig:IsoIntElectrical}e).
Figure~\ref{fig:IsoIntElectrical}f shows, for the present work, the charge density due to screening.  For physical consistency, $\vec{E}$ and $\phi$ must be continuous at the interface, so that HZ and JS must have both an infinitesimally thin charge layer and an infinitesimally thin dipole layer at the interface.

We conclude that outside of the charge screening length  $\ell_{Q}$ (which is very short for metals), the present work and HZ are equally valid for calculating electrical potential, field, and charge, but JS differs significantly.

\begin{figure}[h!]
\centerline{\includegraphics[width=0.46\textwidth]{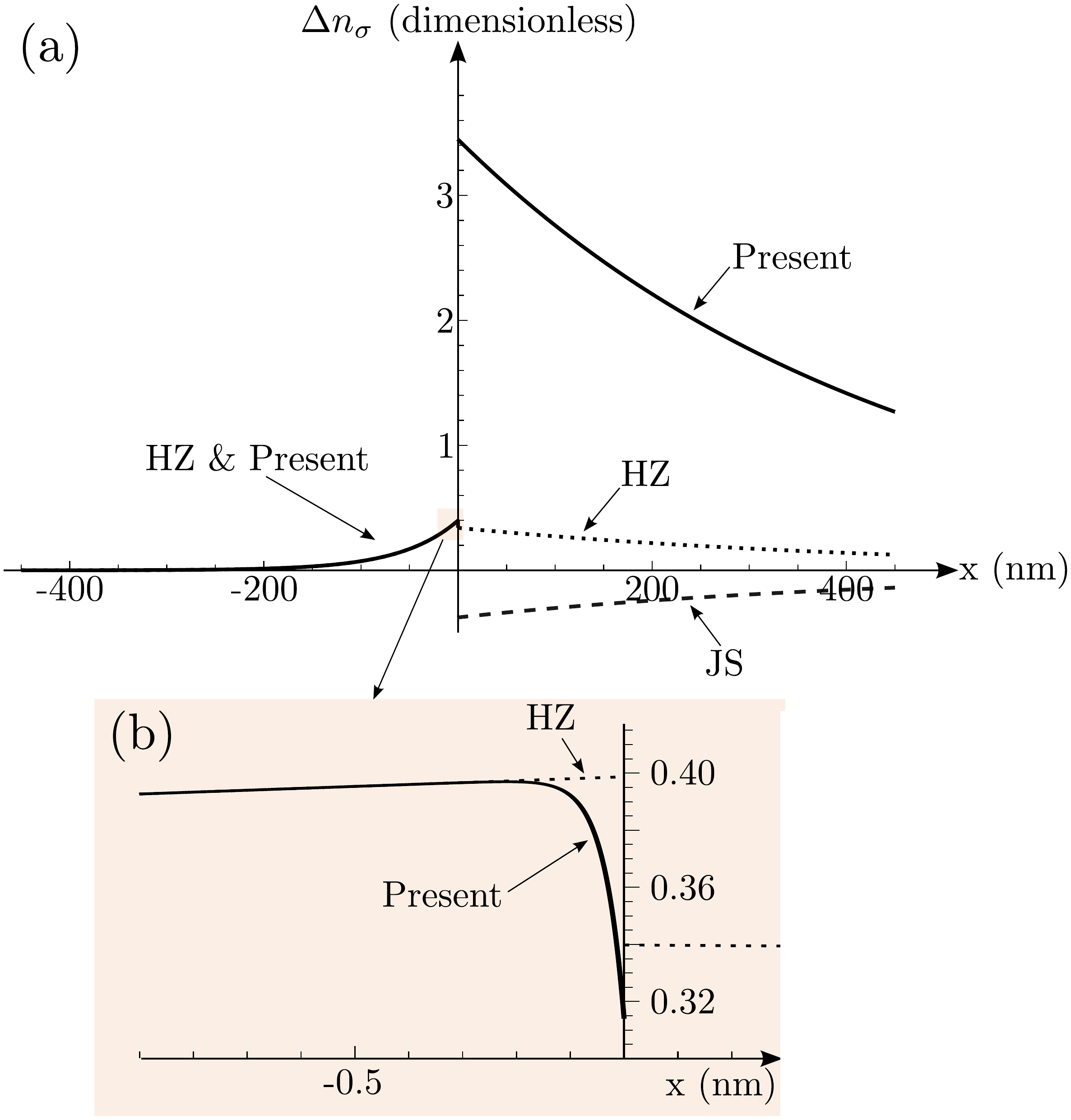}}
\caption[The dimensionless spin accumulation near an interface between copper and cobalt.]
{The dimensionless spin accumulation, given by Eqs.~\eqref{FMspinaccTot} and \eqref{NMspinaccTot}, in arbitrary units near an interface between cobalt ($x<0$) and copper ($x>0$) vs. $x$ (in nm).  
The spin accumulation is shown within approximately (a) 1~$\mu$m and (b) 1~nm of the interface. 
Solid line -- present work; dashed line -- JS; dotted line -- HZ.  
In the FM ($x<0$), HZ nearly coincides with the present work (deviating only within the charge-screening length of the interface, see inset), and the JS-predicted spin accumulation is several orders of magnitude larger and not shown.  In the NM, neither approximation predicts a spin accumulation similar to the present work.}
\label{fig:IsoIntSpin} 
\end{figure}

Figure~\ref{fig:IsoIntSpin} shows  the spin accumulation for the present work, HZ, and JS.  In the non-magnetic material ($x>0$), as shown analytically in Eqs.~\eqref{zetaHZ} and \eqref{zetaJS}, Fig.~\ref{fig:IsoIntSpin}a shows that both HZ and JS differ from the present work by an order of magnitude, with JS having the opposite sign.  Fig.~\ref{fig:IsoIntSpin}b shows that the spin accumulation in the ferromagnet ($x<0$) differs for the present work and HZ; outside of this length, Fig~\ref{fig:IsoIntSpin}a shows that they coincide.  However, the spin accumulation for JS is \textit{six orders of magnitude larger} (and not shown).  This is because JS, by assuming that $\partial \mu_{\uparrow,\downarrow}/\partial n_{\uparrow,\downarrow}=0$, effectively takes $N_{\uparrow,\downarrow} \rightarrow \infty$ so that $N_{\uparrow,\downarrow} \gg N_{\chi}$, whereas Tables~\ref{tab:Numeric1} and \ref{tab:Numeric2} show that the opposite is true for cobalt.

\section{Summary \& Conclusion}
\label{sec:Summary}

Using irreversible thermodynamics, we predict the spin accumulation at an interface between two materials when electric current is driven across the interface.  Although we have numerically studied a FM/NM interface, the theory also applies to FM/FM and NM/NM interfaces.  

We find that both the chemical potentials and the effective magnetic field must be included to predict the spin accumulation in a non-magnetic material -- in fact, for Cu the spin accumulation changes by an order of magnitude on neglect of either contribution.  However, for ferromagnets neglecting the effective magnetic field may be appropriate -- numerically the results are essentially unchanged for Co near a Co/Cu interface.

By including the screening surface mode neglected in previous works, we find an additional term in the spin accumulation for ferromagnets.  For Co near a Co/Cu interface, this term decreases the spin accumulation by $\sim 10 \%$ within a charge-screening length of the interface.  Although this length is on the order of 1-10 \AA~for metals (a length scale negligible in the present macroscopic theory), for ferromagnetic semiconductors this length scale should be much larger.  Note that spin injection from a ferromagnetic semiconductor into a non-magnetic material has been observed by Refs.~\onlinecite{FiederlingNature} and \onlinecite{OhnoAwschalom99}.  To test this spin accumulation due to screening, one may apply a small current to an interface between, say, Ga(Mn)As and Cu.  Using the magneto-optical Kerr effect (MOKE), one may measure the magnetization (and spin-polarization) at the surface.  We expect there to be two nonequilibrium magnetization contributions near the surface, one that decays over the spin-diffusion length $\ell_{\rm sf}$ and one that decays over the screening length $\ell_{\rm Q}$ associated with screening.  The latter effect should be more prominent in ferromagnetic semiconductors.

\section{Acknowledgements}
We would like to acknowledge the support of the Department of Energy through through grant DE-FG02-06ER46278.

\bibliographystyle{apsrev4-1}
\bibliography{IrrThBib}{}

\appendix

\section{Details of the Spin Mode}
\label{appendix:SpinMode}

The details of the solution for the Spin Mode, whose results are presented in Sec.~\ref{subsec:SMode}, are now given.

 Equations~\eqref{muup}-\eqref{mudown} give, with $\delta$ denoting deviations from equilibrium,
\begin{gather}
 \partial_i^2 \delta \bar{\mu}_{{\uparrow}_S} =  \frac{1}{\ell_{{\uparrow}_S}^2} \left(\delta\bar{\mu}_{{\uparrow}_S} - \delta\bar{\mu}_{{\downarrow}_S} \right),\label{muup2}\\
\partial_i^2 \delta\bar{\mu}_{{\downarrow}_S} = -  \frac{1}{\ell_{{\downarrow}_S}^2} \left(\delta\bar{\mu}_{{\uparrow}_S} - \delta\bar{\mu}_{{\downarrow}_S} \right). \label{mudown2}
\end{gather}
Subtracting Eq.~\eqref{mudown2} from Eq.~\eqref{muup2} gives
\begin{align}
\partial_i^2 \left( \delta\bar{\mu}_{{\uparrow}_S} - \delta\bar{\mu}_{{\downarrow}_S} \right) =  \frac{1}{\ell_{\rm{{\rm{sf}}}}^2} \left( \delta\bar{\mu}_{{\uparrow}_S} - \delta\bar{\mu}_{{\downarrow}_S} \right),
\label{mudiffS}
\end{align}
where $\ell_{sf}$ is defined by Eq.~\eqref{lsf}.  
On neglecting $\delta \vec{H}^* \cdot \hat{M}$ and making the identification $\alpha \rightarrow (N_{\uparrow}/\tau_{\uparrow \downarrow}) = (N_{\downarrow}/\tau_{\downarrow \uparrow})$, Eq.~\eqref{lsf} agrees with Ref.~\onlinecite{Hershfield}. 
We use Eqs.~\eqref{lsfupdown} and \eqref{lsf} to find $\alpha$, $\ell_{{\uparrow}_S}$ and $\ell_{{\downarrow}_S}$ in terms of $\ell_{\rm{{\rm{sf}}}}$, $\sigma_{\uparrow}$, and $\sigma_{\downarrow}$, since they are, in principle, measurable:
\begin{gather}
\alpha = \frac{\sigma_{\uparrow} \sigma_{\downarrow}}{e^2 \left(\sigma_{\uparrow}+ \sigma_{\downarrow}\right)\ell_{\rm{{\rm{sf}}}}^2},\label{alpha}\\[1ex]
\ell_{{\uparrow}_S} =  \ell_{\rm{{\rm{sf}}}}\sqrt{\frac{\sigma_\uparrow + \sigma_{\downarrow}}{\sigma_{\downarrow}}}, \quad \ell_{{\downarrow}_S} =  \ell_{\rm{{\rm{sf}}}}\sqrt{\frac{\sigma_\uparrow + \sigma_{\downarrow}}{\sigma_{\uparrow}}}.
\end{gather}

Solving Eq.~\eqref{mudiffS} gives
\begin{align}
\delta\bar{\mu}_{{\uparrow}_S} - \delta\bar{\mu}_{{\downarrow}_S} = e V_{0_S} e^{-x/\ell_{\rm{{\rm{sf}}}}}, \label{muDiff1}
\end{align}
where $V_{0_S}$, with units of electric potential, is unknown, to be determined by boundary conditions.  Since Eq.~\eqref{muDiff1} shows the difference in up- and down-spin magnetoelectrochemical potentials to decay over the length $\ell_{\rm{{\rm{sf}}}}$ from an interface -- this length is called the ``spin-flip" or ``spin-diffusion" length (and sometimes referred to as the ``SDL'').  The length $\ell_{\rm{{\rm{sf}}}}$ may be measurable by employing the Magneto-Optical Kerr Effect\cite{Kerr1877,PostavaFert96} or the Inverse Spin Hall Effect,\cite{ISHE} or may be derived using GMR measurements and theory.\cite{BassHolodyPratt94}


Substitution of Eq.~\eqref{muDiff1} into Eqs.~\eqref{muup2} and \eqref{mudown2} yields
\begin{align}
&\delta\bar{\mu}_{{\uparrow}_S} = \frac{ \ell_{\rm{{\rm{sf}}}}^2}{ \ell_{{\uparrow}_S}^2} e V_{0_S} e^{- x/\ell_{\rm{{\rm{sf}}}}} = \frac{\sigma_{\downarrow}}{\sigma_{\uparrow} + \sigma_{\downarrow}}e V_{0_S} e^{- x/\ell_{\rm{{\rm{sf}}}}},\label{muupS}\\
&\delta\bar{\mu}_{{\downarrow}_S} = - \frac{ \ell_{\rm{{\rm{sf}}}}^2}{ \ell_{{\downarrow}_S}^2} e V_{0_S} e^{- x/\ell_{\rm{{\rm{sf}}}}} = -\frac{\sigma_{\uparrow}}{\sigma_{\uparrow} + \sigma_{\downarrow}}e V_{0_S} e^{- x/\ell_{\rm{{\rm{sf}}}}}. \label{mudownS}
\end{align}
Equations~\eqref{muupS} and \eqref{mudownS} give $\delta\bar{\mu}_{{\uparrow}_S} = -(\ell_{{\downarrow}_S}^2/\ell_{{\uparrow}_S}^2) \delta\bar{\mu}_{{\downarrow}_S} = -(\sigma_{\downarrow}/\sigma_{\uparrow}) \delta\bar{\mu}_{{\downarrow}_S}$, which agrees with Ref.~\onlinecite{Hershfield}.  Substitution of Eqs.~\eqref{muupS} and \eqref{mudownS} into Eq.~\eqref{Fluxes} gives the up- and down- spin carrier currents of Eq.~\eqref{FluxesS}.

We can now can write two independent relations between $\delta n_{{\uparrow}_S}$, $\delta n_{{\downarrow}_S}$, and $\delta \phi_S$.  Equations~\eqref{muDiff1} and \eqref{mubarLin} give the difference of the spin potentials to be
\begin{align}
\delta \bar{\mu}_{{\uparrow}_S}- \delta \bar{\mu}_{{\downarrow}_S} =& \left( \frac{N_{\uparrow} +  N_{\chi}}{N_{\chi}N_{\uparrow}} \right) \delta n_{{\uparrow}_S} - \left(\frac{N_{\downarrow}+ N_{\chi}}{N_{\chi} N_{\downarrow}} \right) \delta n_{{\downarrow}_S} \notag\\
=& e V_{0_S} e^{-x/\ell_{\rm{{\rm{sf}}}}},
\label{muDiff2}
\end{align}
and Eqs.~\eqref{muupS}, \eqref{mudownS}, and \eqref{mubarLin} give the sum of the spin potentials to be
\begin{align}
\delta \bar{\mu}_{{\uparrow}_S} + \delta \bar{\mu}_{{\downarrow}_S} =& \frac{\delta n_{{\uparrow}_S}}{N_{\uparrow}}   + \frac{\delta n_{{\downarrow}_S}}{N_{\downarrow}}   -2 e \delta \phi_S = {C} e V_{0_S} e^{-x/\ell_{\rm{{\rm{sf}}}}}.
\label{muSum1}
\end{align}
In conjunction with Gauss's Law, Eqs.~\eqref{muDiff2} and \eqref{muSum1} give the concentrations and electrical potential in the spin mode. 
Specifically,
we use Eq.~\eqref{muDiff2} to relate $ \delta n_{{\uparrow}_S}$ to $\delta n_{{\downarrow}_S}$, then use Eq.~\eqref{muSum1} to relate $ \delta n_{{\uparrow}_S}$ to $ \delta \phi_S$.  Thus Eq.~\eqref{Gauss} can be written in terms of only $\delta \phi_S$, which we solve. 

Equation~\eqref{muDiff2} gives
\begin{align}
\delta n_{{\downarrow}_S} = &\left[\frac{N_{\downarrow} \left(N_{\uparrow} + N_{\chi} \right)}{N_{\uparrow} \left(N_{\downarrow} + N_{\chi}\right)} \right] \delta n_{{\uparrow}_S} - \left(\frac{N_{\chi} N_{\downarrow} e V_{0_S} }{N_{\downarrow} + N_{\chi}}\right) e^{-x/\ell_{\rm{sf}}}.
\label{ndownTOnup}
\end{align}
Substituting Eq.~\eqref{ndownTOnup} into Eq.~\eqref{muSum1} multiplied by $(N_{\uparrow}/N_{\alpha})(N_{\downarrow} + N_{\chi})$ 
gives
\begin{align}
\delta n_{{\uparrow}_S} =& \frac{N_{\uparrow}}{N_{\alpha}}e V_{0_S} e^{-x/\ell_{\rm{sf}}} \left[N_{\chi} + {C} \left(N_{\downarrow} + N_{\chi} \right) \right]
 \notag\\
 &+ 2 e \frac{N_{\uparrow}}{N_{\alpha}} \left(N_{\downarrow} + N_{\chi}\right)  \delta \phi_S.
\label{nupTOphi}
\end{align}
Substitution of Eqs.~\eqref{ndownTOnup} and \eqref{nupTOphi} into Eq.~\eqref{Gauss} gives
\begin{align}
&\partial_x^2 \delta \phi_S  = \frac{2  N_{\beta}^{2}}{N_S N_{\alpha} \ell_{\rm{sf}}^2} \notag\\
&\times \left\{ \delta \phi_S + \frac{V_{0_S}}{2}\left[  \frac{N_{\chi}}{N_{\beta}^2} \left(N_{\uparrow}  - N_{\downarrow}  \right) + {C} \right] e^{-x/\ell_{\rm{sf}}} \right\}.
\label{Gauss4}
\end{align}
The solution for $\delta \phi_{S}$ is given above as Eq.~\eqref{phiS}.

Substituting Eq.~\eqref{phiS} into Eqs.~\eqref{nupTOphi} and \eqref{ndownTOnup} gives the up- and down- spin concentrations of Eqs.~\eqref{nupS} and \eqref{ndownS}. 
Thus, the charge distribution $\delta \rho = -e \left( \delta n_{{\uparrow}} + \delta n_{{\downarrow}} \right) $ associated with the spin mode is
\begin{align}
\delta \rho_S 
= \left[ \frac{N_{\chi} \left(N_{\uparrow} - N_{\downarrow} \right) + {C} N_{\beta}^2  }{N_S N_{\alpha} - 2 N_{\beta}^2} \right] N_{S} e^2  V_{0_S} e^{-x/\ell_{\rm{sf}}} 
, \label{rhoS}
\end{align}
which is nonzero in a ferromagnet.  (This result can also be obtained by using Eq.~\eqref{phiS} and Gauss's Law.)  Further, subtraction of Eq.~\eqref{ndownS} from Eq.~\eqref{nupS} yields the spin accumulation of Eq.~\eqref{SFMspinacc}.

\section{Boundary Conditions for Current Crossing an Isolated Interface}
\label{appendix:IsolatedInt}
Boundary conditions {\bf (i-viii)} for an isolated interface (that is, one that is effectively an infinite distance from any other interface) through which an electric current is passed are discussed in Sec.~\ref{sec:IsoInt}.  They are here found explicitly, in numerical order.

Conditions {\bf (i-ii)}: From Eqs.~\eqref{PhiTot} and \eqref{ETot}, continuity of $\delta \phi$ and $\delta E$ across the interface at $x_{\rm{int}}=0$ gives
\begin{gather}
\xi^{\rm (I)} {V_{0_S}^{\rm (I)}} + {V_{0_Q}^{\rm (I)}}  + V_{0_{dc}}^{\rm (I)} = \xi^{\rm{(II)}} {V_{0_S}^{\rm{(II)}}} + {V_{0_Q}^{\rm{(II)}}} +V_{0_{dc}}^{\rm{(II)}}, \label{phiI2}\\[1ex]
-\frac{\xi^{\rm (I)} V_{0_S}^{\rm (I)}}{{\ell_{\rm{sf}}^{\rm (I)}}} - \frac{V_{0_Q}^{\rm (I)}}{\ell_Q^{\rm (I)}}  + E_{0_{dc}}^{\rm (I)}= \frac{\xi^{\rm{(II)}} V_{0_S}^{\rm{(II)}}}{\ell_S^{\rm{(II)}}} + \frac{V_{0_Q}^{\rm{(II)}}}{\ell_Q^{\rm{(II)}}}   + E_{0_{dc}}^{\rm{(II)}}. \label{EI2}
\end{gather}
Recall that $\xi =0$ for a non-magnetic material.

Condition {\bf (iii)}: From Eq.~\eqref{JTot}, continuity of the electric current across the interface gives
\begin{gather}
\left(\sigma_{\uparrow}^{\rm (I)} + \sigma_{\downarrow}^{\rm (I)}\right) E_{0_{dc}}^{\rm (I)} = \left(\sigma_{\uparrow}^{\rm (II)} + \sigma_{\downarrow}^{\rm (II)}\right) E_{0_{dc}}^{\rm (II)}.
\label{JI2}
\end{gather}

Condition {\bf (iv)}:  Although the electric current is continuous everywhere, in principle at the interface there may be spin scattering, so that spin current is not continuous across the interface.  However, we neglect interfacial spin scattering (as is typical in this type of theory).  We thus take
\begin{align}
j_{\uparrow}^{\rm (I)} (0) \equiv j_{\uparrow}^{\rm{(II)}} (0) ,\qquad
j_{\downarrow}^{\rm (I)} (0) \equiv j_{\downarrow}^{\rm{(II)}} (0) .
\label{JIcontinuous}
\end{align}
Using Eq.~\eqref{jupTot}, the first of these can be written as
\begin{align}
&- \sigma_{\uparrow}^{\rm (I)}  E_{0_{dc}}^{\rm (I)} + \left(\frac{\sigma_{\uparrow}^{\rm (I)} \sigma_{\downarrow}^{\rm (I)}}{\sigma_{\uparrow}^{\rm (I)} + \sigma_{\downarrow}^{\rm (I)}} \right) \frac{V_{0{_S}}^{\rm (I)}}{\ell_{\rm sf}^{\rm (I)}} \notag\\
&= 
- \sigma_{\uparrow}^{\rm{(II)}}  E_{0_{dc}}^{\rm{(II)}} + \left(\frac{\sigma_{\uparrow}^{\rm{(II)}} \sigma_{\downarrow}^{\rm{(II)}}}{\sigma_{\uparrow}^{\rm{(II)}} + \sigma_{\downarrow}^{\rm{(II)}}} \right) \frac{V_{0{_S}}^{\rm{(II)}}}{\ell_{\rm sf}^{\rm{(II)}}} .
\label{JupIcont}
\end{align}
As discussed above, the second relation given in Eq.~\eqref{JIcontinuous} is then automatically satisfied by condition {\bf (iii)}, which constrain the sums of the up- and down-spin currents.

Conditions {\bf (v-vi)}: 
The spin currents across the interface are given by\cite{JohnsonSilsbee,SearsSasHeating} 
\begin{gather}
{j_{\uparrow}}_{\rm{int}} 
= - \frac{g_{\uparrow}}{e^2} (\Delta \bar{\mu}_{\uparrow})_{\rm{int}},\label{JupI}\\
{j_{\downarrow}}_{\rm{int}} 
= - \frac{g_{\downarrow}}{e^2} (\Delta \bar{\mu}_{\downarrow})_{\rm{int}}, \label{JdownI}
\end{gather}
were, $(\Delta)_{\rm int}$ denotes the difference between the value just on the right of the interface ($x \rightarrow 0^{+}$) and the value just on the left ($x \rightarrow 0^{-}$).  
Since without the electric field associated with the $dc$ mode there is no steady-state current, the currents are proportional to the differences in $\delta \bar{\mu}$ rather than $\bar{\mu}$.  We now find $(\Delta \delta \bar{\mu})_{\rm{int}}$ for each mode and then substitute them into Eqs.~\eqref{JupI}-\eqref{JdownI}.

The charge mode has $\delta \bar{\mu}_{{\uparrow}_Q} = 0 = \delta \bar{\mu}_{{\downarrow}_Q}$, so by Eqs.~\eqref{JupI}-\eqref{JdownI} it does not affect the current crossing the boundary.  At the $x=x_{\rm{int}}=0$ interface, Eqs.~\eqref{muupS} and \eqref{mudownS} give 
\begin{align}
&(\Delta \delta \bar{\mu}_{{\uparrow}_S})_{\rm{int}} \notag\\
&= \left(\frac{\sigma_{\downarrow}^{\rm{(II)}}}{\sigma_{\uparrow}^{\rm{(II)}}+ \sigma_{\downarrow}^{\rm{(II)}}}\right) e V_{0_S}^{\rm{(II)}} - \left( \frac{\sigma_{\downarrow}^{\rm (I)}}{\sigma_{\uparrow}^{\rm (I)}+ {\sigma_\downarrow^{\rm (I)}}} \right) e V_{0_S}^{\rm (I)} ,
\label{deltamuupS}\\[1ex]
&(\Delta \delta \bar{\mu}_{{\downarrow}_{S}})_{\rm{int}} \notag\\
&= -\left( \frac{\sigma_{\uparrow}^{\rm{(II)}}}{\sigma_{\uparrow}^{\rm{(II)}}+ \sigma_{\downarrow}^{\rm{(II)}}} \right) e V_{0_S}^{\rm{(II)}}+ \left( \frac{\sigma_{\uparrow}^{\rm (I)}}{\sigma_{\uparrow}^{\rm (I)}+ \sigma_{\downarrow}^{\rm (I)}} \right) e V_{0_S}^{\rm (I)}.
\label{deltamudownS}
\end{align}
At the interface, Eq.~\eqref{phidc} gives
\begin{align}
(\Delta \delta \bar{\mu}_{{\uparrow}_{dc}} )_{\rm{int}} = (\Delta \delta \bar{\mu}_{{\downarrow}_{dc}})_{\rm{int}} =& -e \left( V_{0_{dc}}^{\rm{(II)}} - V_{0_{dc}}^{\rm (I)} \right).
\label{deltamudc}
\end{align}

Substitution of Eqs.~\eqref{deltamuupS}-\eqref{deltamudc} into Eqs.~\eqref{JupI}-\eqref{JdownI} yields
\begin{align}
&{j_{\uparrow}}_{\rm{int}} = - \frac{g_{\uparrow}}{e} \left[\frac{\sigma_{\downarrow}^{ \rm{(II)}} V_{0_S}^{\rm{(II)}}}{\sigma_{\uparrow}^{\rm{(II)}}+ \sigma_{\downarrow}^{ \rm{(II)}}}  -\frac{\sigma_{\downarrow}^{\rm (I)} V_{0_S}^{\rm (I)}}{\sigma_{\uparrow}^{\rm (I)}+ \sigma_{\downarrow}^{\rm (I)}} - \left( V_{0_{dc}}^{\rm{(II)}} - V_{0_{dc}}^{\rm (I)} \right)\right],\label{JupI2}\\
&{j_{\downarrow}}_{\rm{int}}= - \frac{g_{\downarrow}}{e} \left[-\frac{\sigma_{\uparrow}^{\rm{(II)}} V_{0_S}^{\rm{(II)}}}{\sigma_{\uparrow}^{ \rm{(II)}}+ \sigma_{\downarrow}^{\rm{(II)}}} +\frac{\sigma_{\uparrow}^{\rm (I)} V_{0_S}^{\rm (I)}}{\sigma_{\uparrow}^{\rm (I)}+ \sigma_{\downarrow}^{\rm (I)}}  - \left( V_{0_{dc}}^{\rm{(II)}} - V_{0_{dc}}^{\rm (I)} \right)\right]. \label{JdownI2}
\end{align}
We take 
\begin{equation}
{j_{\uparrow}}_{\rm{int}} \equiv j_{\uparrow}^{\rm{(II)}} (0), \qquad {j_{\downarrow}}_{\rm{int}} \equiv j_{\downarrow}^{ \rm{(II)}} (0).
\end{equation}  
By Eq.~\eqref{JIcontinuous} one may equivalently use ${j_{\uparrow}}_{\rm{int}} \equiv j_{\uparrow}^{\rm (I)} (0)$ and ${j_{\uparrow}}_{\rm{int}} \equiv j_{\downarrow}^{\rm (I)} (0)$.  Respective substitution of Eqs.~\eqref{jupTot} and \eqref{jdownTot} 
into Eqs.~\eqref{JupI2} and \eqref{JdownI2} 
gives
\begin{align}
&j_{\uparrow}^{\rm{(II)}} (0) =  -  \frac{\sigma^{\rm{(II)}}_{\uparrow}}{e} E_{0_{dc}}^{\rm{(II)}}
+\left(\frac{\sigma_{\uparrow}^{\rm{(II)}} \sigma_{\downarrow}^{ \rm{(II)}}}{\sigma_{\uparrow}^{\rm{(II)}} + \sigma_{\downarrow}^{ \rm{(II)}}} \right) 
\frac{V_{0_{S}}^{\rm (II)}}{e\ell_{\rm{sf}}^{ \rm{(II)}}} \notag\\
&= - \frac{g_{\uparrow}}{e} \left[\frac{\sigma_{\downarrow}^{\rm{(II)}} V_{0_S}^{\rm{(II)}}}{\sigma_{\uparrow}^{\rm{(II)}}+ \sigma_{\downarrow}^{\rm{(II)}}}  -\frac{\sigma_{\downarrow}^{\rm (I)} V_{0_S}^{\rm (I)}}{\sigma_{\uparrow}^{\rm (I)}+ \sigma_{\downarrow}^{\rm (I)}} - \left( V_{0_{dc}}^{\rm{(II)}} - V_{0_{dc}}^{\rm (I)} \right)\right],\label{JupI3}\\
&j_{\downarrow}^{\rm{(II)}} (0) = 
 -  \frac{\sigma^{\rm{(II)}}_{\downarrow}}{e} E_{0_{dc}}^{\rm{(II)}}
- \left(\frac{\sigma_{\uparrow}^{\rm{(II)}} \sigma_{\downarrow}^{\rm{(II)}}}{\sigma_{\uparrow}^{\rm{(II)}} + \sigma_{\downarrow}^{\rm{(II)}}} \right)
\frac{V_{0_S}^{\rm{(II)}} }{e\ell_{\rm{sf}}^{\rm{(II)}}}\notag\\
&= - \frac{g_{\downarrow}}{e} \left[-\frac{\sigma_{\uparrow}^{\rm{(II)}} V_{0_S}^{\rm{(II)}}}{\sigma_{\uparrow}^{\rm{(II)}}+ \sigma_{\downarrow}^{\rm{(II)}}} +\frac{\sigma_{\uparrow}^{\rm (I)} V_{0_S}^{\rm (I)}}{\sigma_{\uparrow}^{\rm (I)}+ \sigma_{\downarrow}^{\rm (I)}}  - \left( V_{0_{dc}}^{\rm{(II)}} - V_{0_{dc}}^{\rm (I)} \right)\right]. \label{JdownI3}
\end{align}

Condition {\bf (vii)}: The total electric current $J_{app}$ is known, so the total electric current in material I can be written using Eq.~\eqref{JTot} as
\begin{align}
- \left( \sigma_{\uparrow}^{\rm (II)} + \sigma_{\downarrow}^{\rm (II)} \right) E_{0_{dc}}^{\rm (II)} = J_{app}.
\label{JB2}
\end{align}
Equation~\eqref{JI2} then guarantees that the total current in material I also equals $J_{app}$.

Condition {\bf (viii)}: There is an arbitrary constant potential.  We set
\begin{align}
V_{0_{dc}}^{\rm{(II)}} \equiv 0.
\label{ArbPhi2}
\end{align}

The eight conditions explicitly given by Eqs.~\eqref{phiI2}-\eqref{JI2}, \eqref{JupIcont}, and \eqref{JupI3}-\eqref{ArbPhi2} are general for an isolated interface between any two materials I and II, and their solution gives the eight unknowns $E_{0_{dc}}^{\rm (I,II)}$, $V_{0_{dc}}^{\rm (I,II)}$, $V_{0{_Q}}^{\rm (I,II)}$, and $V_{0{_S}}^{\rm (I,II)}$.  

\end{document}